# Towards Computational UIP in Cubical Agda

MPRI M2 Thesis by Yee-Jian Tan

Study Programme: Foundations of Computer Science

Ecole Polytechnique, Institut Polytechnique de Paris

Leuven, Belgium,  25 November 2025
(improved with incorporated feedback)

supervised by
Prof. Dr. Dominique Devriese
Dr. Andreas Nuyts

# Abstract

Cubical Agda [VMA21] is a proof assistant that implements a flavour of Homotopy Type Theory (HoTT) [Uni13] called Cubical Type Theory [Coh+17]. Some advantages of Cubical Type Theory over intensional Martin-Löf Type Theory include Quotient Inductive Types (QITs), which exist as instances of Higher Inductive Types, and functional extensionality, which is provable in Cubical Type Theory [Coh+17]. However, HoTT features an infinite hierarchy of equalities that may become unwieldy in formalisations. Fortunately, QITs and functional extensionality are both preserved even if the equality levels of Cubical Type Theory are truncated to only homotopical Sets (h-Sets) [Uni13]. In other words, removing the univalence axiom from Cubical Type Theory and instead postulating a conflicting axiom: the Uniqueness of Identity Proofs (UIP) postulate. Since univalence is proved in Cubical Type Theory from the so-called Glue Types, therefore, it is known that one can first remove the Glue Types (thus removing univalence) and then set-truncate all equalities (essentially assuming UIP), à la XTT [SAG22]. The result is a "h-Set Cubical Type Theory" that retains features such as functional extensionality and QITs.

However, in Cubical Agda, there are currently only two unsatisfying ways to achieve h-Set Cubical Type Theory. The first is to give up on the canonicity of the theory and simply postulate the UIP axiom, while the second way is to use a standard result stating "type formers preserve h-levels" to manually prove UIP for *every* defined type. The latter is, however, laborious work best suited for an automatic implementation by the proof assistant. In this project, we analyse formulations of UIP and detail their computation rules for Cubical Agda, and evaluate their suitability for implementation. We also implement a variant of Cubical Agda without Glue, which is already compatible with postulated UIP, in anticipation of a future implementation of UIP in Cubical Agda.

# Contents

# Chapter 1

# Introduction

In Dependent Type Theories such as Martin-Löf Type Theory (MLTT), equality is encoded as an intensional identity type with only the reflexivity constructor. Well-known properties about equality, such as symmetry, transitivity, congruence[1], and substitutivity[2] can all be derived from its elimination principle, $J$ [Uni13], but whether the identity type must be uniquely inhabited was not clear. The Uniqueness of Identity Proofs (UIP) postulate which states that proofs of equality, if they existed, are unique, proposes a positive answer to the question above, and is supported by the set model of type theory [Hof97]. In 1994, Hofmann and Streicher constructed the groupoid model of type theory that refutes the UIP postulate [HS94], thus showing that UIP does not hold in all models of type theory, and hence can neither be proven nor disproved within MLTT. This paved the way for Homotopy Type Theory (HoTT) [Uni13], whose philosophy, as encapsulated by the univalence axiom, is that proofs of equality are not necessarily unique. The univalence axiom states roughly that:

$$\text{For any types } A, B : \text{Type}, \, (A =_{\text{Type}} B) \cong (A \cong B).$$

In particular, an implication is that the proofs of equality between $A$ and $B$ are in bijection with the equivalences[3] between $A$ and $B$. Since there are types that are equivalent in more than one way, it follows that corresponding equality proofs are distinct. For example, there are two equivalences between the types Bool and Bool, namely the identity function and the "not" operator, resulting in two distinct proofs of equality Bool $=_{\text{Type}}$ Bool, directly contradicting UIP. By consequence, the univalence axiom and the UIP postulate are incompatible. Therefore, type theories can choose to adopt at most one of univalence and UIP, but certainly not both.

Cubical Agda [VMA21] implements Cubical Type Theory [Coh+17], a flavour of HoTT, and thus also has some of the advantages of HoTT: functional extensionality not only holds, but is provable in Cubical Type Theory, and Quotient Inductive Types (QITs) are available as a particular case of Higher Inductive Types (HITs). As with other HoTTs, Cubical Type Theory has univalence and is thus incompatible with UIP. At first glance, the title of this report, "Towards Computational UIP in Cubical Agda" might seem like doing the impossible. However, in Cubical Type Theory, a specific component, called the Glue types, is responsible for the provability of the univalence axiom. Thus, as a special case, one can safely postulate UIP in a Cubical Type Theory *without* Glue types, truncating away the higher equalities while remaining consistent. This results in a type system where types are so-called "$h$-Sets" with only one equality level, yet keeps several desirable features of HoTT such as functional extensionality and QITs. As a result, researchers and Cubical Agda users have been curious about a Cubical Type Theory compatible with UIP [Dan19, Shu17] as well as an actual implementation as a Cubical Agda variant that they can work in.

Although attractive as a theory, merely *postulating* UIP in a proof assistant like Cubical Agda (albeit without Glue) blocks computation and removes *canonicity* (e.g. normal terms of inductive types start with constructors in the empty context) from the system. Instead of invoking the UIP postulate at every type and resulting in computation-blocking terms, an alternative is to manually apply standard results of HoTT, proving UIP inductively based on type derivations through type formers, as long as the base types have the UIP property and Higher Inductive Types have a UIP "set-truncating" constructor – essentially a proof by induction on type formation. That said, deriving these UIP proofs manually for *every* type is laborious and tedious work that should be taken care of by the proof assistant. This project aims to fill this gap by allowing automatic proofs of the UIP property on arbitrary types recursively based on their type formers, just as a user would have manually written.

---

[1] congruence: $x = y \Rightarrow \forall f, fx = fy$
[2] substitutivity: $x = y \Rightarrow \forall P, Px \Rightarrow Py$
[3] Two types $A$ and $B$ are equivalent if there exists an invertible function $f : A \to B$ between them.

The challenge of the project is twofold: other than the technical difficulty of extending the (Cubical) Agda implementation to be compatible with and compute UIP, theoretically, we also have to select a sufficiently general formulation of UIP suitable for implementation and carefully state its computation rule for each type former. Essentially, these computation rules act as the inductive cases in an induction on type derivation to show UIP for all types. For example, the UIP computation rule for the Sigma type former is a proof of the following statement:

> A Sigma type $\Sigma_{a:A} B(a)$ has the UIP property if the type $A$ and the types $B(a)$ for every $a : A$ all have the UIP property.

Ideally, these proofs to be turned into computation rules should be as simple as possible to result in an elegant theory, which turned out to have no straightforward answer. In this report, we consider an even more general (yet still equivalent) statement of UIP, called the square-filling property in order to "strengthen (usability-wise) the induction hypothesis" for the proof by induction on type formers. We detail the computation rules for two versions of the square-filling property: the homogeneous square-filling property SqFill for types, and the heterogeneous square-filling Property SqPFill for (potentially higher) squares of types, forming a basis for their implementation.

The contributions of this internship are as follows:
- adding an Agda Cubical variant [Tan25a] in which Glue types are disabled, and
- detailing the UIP computational rules for a Cubical Type Theory without glue for the type formers
  ‣ Pi (dependent functions),
  ‣ Sigma (dependent pairs),
  ‣ Coproducts, and
  ‣ Path types

  in two versions of the square-filling property: the homogeneous SqFill and the heterogeneous SqPFill. Their proofs are provided in Cubical Agda (https://swampertx.github.io/hset-cubical/Everything.html); more precisely, in the Glue-free variant which we implemented.

The rest of the report is organised as follows: Chapter 2 introduces the underlying concepts such as HoTT, Cubical Type Theory, and h-levels. Chapter 3 explains the "Agda Cubical without Glue" implementation as well as preliminary efforts preceding the actual UIP implementation. Chapter 4 explains our two square-filling formulations of UIP along with their computational rules through the four type formers: Pi, Sigma, Coproducts, and Path types. Finally, we investigate related works and future work in Chapter 5.

# Chapter 2

# Primer

This section gives the technical background to Chapter 4 and also reports on the learning done during this internship. We start by assuming only the knowledge in Dependent Type Theories which I started the internship with, such as Martin-Löf Type Theory covered in MPRI 2.7.1 [GLT89, Wer24] or Calculus of Inductive Constructions covered in MPRI 2.7.2 [Coq24, FW24].

## 2.1 The Uniqueness of Identity Proofs (UIP) postulate

In Type Theory, equality between two terms of the same type is expressed as a special type, also known as the Martin-Löf identity type. For example, the identity type $\text{Id}_\mathbb{N}(0, 1)$ between natural numbers 0 and 1, corresponding to the proposition $0 = 1$, is certainly uninhabited since the proposition is false. On the other hand, types corresponding to provable propositions, such as $\text{Id}_\mathbb{N}(0, 0)$ corresponding to $0 = 0$ are inhabited, by virtue of reflexivity of equality. For example, the identity type $\text{Id}_\mathbb{N}(0, 0)$ has an inhabitant $\text{refl} : \text{Id}_\mathbb{N}(0, 0)$, the reflexivity term constructor.

We can further ask: if there is a proof of equality, is that the *only* proof of that equality? Are there more ways that equality can be proven? In other words, is an inhabitant of the identity type always unique? This question, known as the Uniqueness of Identity Proofs (UIP) postulate, was answered negatively in 1994: it is independent of the syntax of type theory [HS94]. This is done by two important models of type theory: Hoffman's set model of type theory [Hof97], in which UIP holds for all types, while Hoffman and Streicher's groupoid model [HS94] refutes UIP.

Thus, type theories could either assume that proofs of equality between terms must be unique, i.e. the UIP assumption, or assume that proofs of equality between terms can be distinct. Homotopy Type Theory grew from the latter viewpoint: types are higher groupoids, in which the identity type of two terms is itself a groupoid, and the identity type of terms in the identity types, too, and so on.

## 2.2 Homotopy Type Theory (HoTT)

On top of the "propositions as types, proofs as programs" correspondence with logic, HoTT further gives a geometrical (more precisely, homotopical) correspondence for type theory: types as spaces (up to homotopy), terms as points in the space, and equality between terms as paths between points. However, Homotopy Type Theory requires two axioms (actually just one, since functional extensionality is derivable from univalence [Uni13]) to form a coherent theory that allows characterising the path types under type formers: paths between functions should behave like functions of paths (functional extensionality), and paths between types should behave like equivalences, i.e. invertible functions (univalence). However, naively postulating axioms discounts the computational advantage of type theory: axioms inhabit a type without any computational behaviour and computations become "stuck". This breaks canonicity, which says, for example, closed terms of Booleans are either equal to true or false. That was the problem faced by "Book HoTT", the theory of HoTT as proposed and proven in the HoTT book [Uni13], and researchers worked on finding a model of type theory in which these axioms have computational meaning, meaning that they could be "assembled" and proven from lower-level primitive computations.

## 2.3 Cubical Type Theory

Such is the contribution of Cubical Type Theory [Coh+17]: a type theory based on cubical sets that gives computational meaning to both functional extensionality and univalence axioms, in the sense that functional extensionality can be proven from the interval type $I$ [Shu11], and univalence from the Glue Types [Coh+17].

### 2.3.1 The Interval Type $I$

**Notation.** We write $\equiv$ to denote intensional/propositional equality; $=$, $\triangleq$, and $:=$ for extensional/definitional equality.

In topology, paths are defined as functions from the unit interval $[0, 1] \subset \mathbb{R}$. Cubical Type Theory takes the "equalities as paths" correspondence literally: equalities in a given type $A$ are functions $I \to A$ from the unit interval type $I$ with two inhabitants (0 and 1, distinct from the natural numbers 0 and suc 0) to $A$. Therefore, given any two inhabitants $a\ b : A$, a proof of equality $p$ from $a$ to $b$, written as $p : a \equiv_A b$, is none other than a function $p : I \to A$ from the unit interval type $I$ to $A$, with endpoints

$$p(0) \triangleq a \text{ and } p(1) \triangleq b.$$

By this literal interpretation, functional extensionality is direct: if two functions $f, g : A \to B$ are pointwise equal, that is, given any point $a : A$ from the domain $A$, there is a path $p_a : I \to B$ between the images $f(a)$ and $g(a)$, then the path $F : I \to (A \to B)$ between the functions $f$ and $g$ can be simply defined by swapping arguments:

$$F(i)(a) := p_a(i),$$

thus we have that $f, g : A \to B$ are equal as functions.

**Heterogeneous Paths.** We can similarly define paths of types such as $A, B : I \to \text{Type}$. This also gives rise to **heterogeneous** path types, such as PathP $A\ a_0\ a_1$, the type of paths between $a_0 : A\ 0$ and $a_1 : A\ 1$, where the ambient type $(\lambda(i : I).A\ i)$ varies along the path. The inhabitants of heterogeneous path types are called heterogeneous paths. On the other hand, a path type is said to be **homogeneous** if the ambient type is constant throughout (such as the ambient type $\mathbb{N}$ in $0 \equiv_\mathbb{N} 1$), and its inhabitants (if they exist) are called homogeneous paths.

**De Morgan Algebra.** Other than $0, 1 : I$, the interval type $I$ is also endowed with the min ($\wedge$), max ($\vee$), and negation ($\sim$) operations which form a De Morgan algebra, i.e. a bounded distributive lattice. Distributive and De Morgan laws hold definitionally, but the laws of excluded middle ($i \vee \sim i = 1$) or absurdity ($i \wedge \sim i = 0$) **do not**. We discuss some consequences and observations in Appendixs A.1 and A.2 due to the nature of De Morgan Algebras being less structured than Boolean algebras, but $I$ being a Boolean algebra leads to problems noticed in [Coq15].

### 2.3.2 Kan Operations and Glue Types

The primitive operations on paths are the so-called "Kan operations", and the only two such primitives in Cubical Agda are

 (i) the homogeneous composition (hcomp), which composes paths together to form a new one, and
(ii) the transport operation (transp), which "transports" objects along path-equivalent types.

Now, the common properties of equalities can all be derived: *reflexivity* by constant functions from $I$, *symmetry* by negating the interval variable of a path, *transitivity* by the hcomp operation, and *substitutivity* by applying transp with predicates.

***n*-cubes.** Composition of paths using hcomp is done for $n$-dimensional cubes of paths. A 0-dimensional cube is a point, 1-dimensional cube is a line (path between points), a 2-dimensional cube is a square (path between paths), and a 3-dimensional cube (path between squares) is, well, a cube. For example, in the 2-dimensional case, hcomp takes an open square without a lid (i.e. three sides of a hollow square) and gives it a lid by composing its sides, resulting in a full hollow square. A filled square can be represented as a path between paths. We will explain more about hcomp and transp below.

From hcomp and transp, many basic operations can be formed, such as
- the **heterogeneous** composition comp, which is the heterogeneous version of hcomp, and
- the homogeneous filling composition hfill, which *fills* a partially-defined $n$-cube.

We leave the technical definitions of these operations to [Coh+17], and give here only the necessary (pictorial) intuition of hcomp, transp, and comp for understanding Chapter 4.

**hcomp.** Informally, in the 2-D case (Figure 1), hcomp states that if we have a hollow square with only the base and (homogeneous) partial sides, hcomp can return a lid that agrees with the partial sides at the endpoints. If the composition is heterogeneous, it is called a comp, and takes an extra argument about the concrete heterogeneous types in question. For example, in the 3-D case (Figure 2), we start with a bottom square $p$ and *heterogeneous* partial sides $u$ that aligns with $p$ at the boundaries and comp will return a top lid that aligns with $u$ at the boundaries.

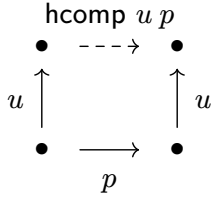
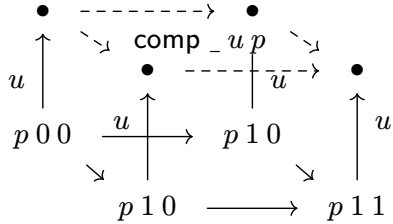
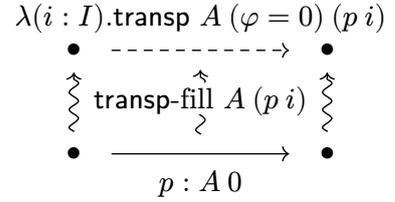

Figure 1: hcomp of a path $p$ along partial sides $u$.

Figure 2: comp of a square $p$ along partial sides $u$.

Figure 3: Transporting a path $p$ along a line of types $\lambda\,(i:I).A\,i$.

To simplify our proofs in Chapter 4, instead of writing out the full details, we will sometimes draw such diagrams of hollow squares or hollow cubes with partial sides, and directly state that the lid will be given by hcomp or comp. We invite interested readers to inspect the full proofs (https://swampertx.github.io/hset-cubical/Everything.html) in Cubical Agda where the boundary conditions are mechanically checked.

**transp.** The transport operation transp can also be understood as a "heterogeneous composition" (*à la* comp) whose partial sides are all "reflexive" (up to equality of types).

More precisely, the transport (transp $A\,\varphi\,a$) operation takes a path of types $A : I \to$ Type, the "transportee" $a : A\,0$, and returns an inhabitant of $A\,1$ which we call the "target". Additionally, it accepts a formula $\varphi : I$ which specifies that at all solutions of $\varphi = 1$, the path of types $A$ is definitionally constant/reflexive. Therefore, when $\varphi = 1$, Cubical Agda will decide that the transportation is definitionally the identity. The special case transport $A\,p := $ transp $A\,0\,p$ where $\varphi$ is specialised to 0 is defined as transp where $A$ is expected to constant when $\varphi := 0 \stackrel{?}{=} 1$, which never holds, thus Cubical Agda never identifies the transportee and the target.

A useful fact about transp is that the transportee and the resulting target are path equivalent. The proof of its equivalence is commonly called the "transp-filler" (Figure 3).

> ℧ **Proposition 2.1.** (transp-fill): If transporting $(a : A\,0)$ along a line of types $(A : I \to $ Type$)$ that stays constant when $\varphi = 1$ results in $b := $ transp $\varphi\,A\,a$, then $a$ and $b$ are equivalent up to a path $p : $ PathP $A\,a\,b$.

*Proof.* The transp-filler $p$ can be defined as another transp itself, by cleverly setting its $\varphi$:

$$p := \lambda(i : I).\ \text{transp}\ [\lambda(j : J).A\,(i \wedge j)]\ (\sim i \vee \varphi)\ a.$$ □

We use squiggly arrows ($\leadsto$) to represent transp-filler paths (Figure 3).

**Irregularity.** A shortcoming of Cubical Type Theory is the lack of "regularity": transport along a reflexive path of types is **not** definitionally the identity function: Cubical Type Theory with regularity might have difficulties showing properties such as canonicity or normalisation [Swa18]. However, it is propositionally equal to the identity function due to the pointwise paths given by Proposition 2.1.

**Glue Types.** The *type* constructor Glue is akin to hcomp as a *path* constructor: Glue takes an $n$-cube of types and partial sides and returns an aligned "lid" $n$-cube of types, where the partial walls are not $n$-cube of types but **equivalences** (i.e. invertible functions).

## 2.4 $h$-levels and Set-Truncation

Even though HoTT allows an arbitrary multiplicity of paths and levels of paths (paths, paths between paths, paths between paths between paths, …), not every type has an infinite level of paths. In particular, Hedberg's Theorem shows that types with decidable equality, such as the booleans $\mathbb{B}$ and natural numbers $\mathbb{N}$, satisfy the UIP property [Uni13]. These types are called "Sets" because they resemble sets where equality is extensional. If one looks at the possible types of paths (the path space) of these set-like types, for example $0 \equiv_\mathbb{N} 0$ and $0 \equiv_\mathbb{N} 1$, they behave like propositions: either empty or uniquely inhabited, hence called "Propositions". In turn, the path space of propositions is contractible: that is, they have a unique inhabitant. The path space of contractible types is again contractible, and we have reached a limit.

This classification of types is called **$h$-levels** ($h$ for homotopy), also called **$n$-types** in the HoTT Book. The Agda Cubical library assigns contractible types an $h$-level of 0, propositional types an $h$-level of 1, sets an $h$-level of 2, and in general, an $h$-level of $(n+1)$ to types whose path spaces are of $h$-level $n$.[4] In other words, the $h$-level of a type $A$ can be understood as "how many levels of non-trivial path spaces of $A$ are there".

### 2.4.1 Higher Inductive Types and Truncation

When defining inductive types in HoTT, not only can we declare constructors for elements, we can also declare equality/path constructors, making their path spaces non-trivial, and recursively on higher path spaces too. The resulting inductive types are called **Higher Inductive Types** (HITs) for their higher equalities. The simplest non-trivial HIT is the Circle $\mathbb{S}^1$, with only one constructor base : Circle, and a non-trivial path loop : base ≡ base, i.e. loop is not equal to refl {x = base}.

Conversely, equality levels of HITs can also be made trivial by adding constructors that contract path spaces. For example, adding the isSet : $\Pi(x\ y : A).\Pi(p\ q : x \equiv y).p \equiv q$ constructor trivialises the path space of $A$ as well as *all* the path spaces "above it", therefore the HIT is also said to be **truncated**. For example, the Circle $\mathbb{S}^1$ truncated by isSet is isomorphic to the unit type. In fact, Quotient Inductive Types are a special case of HITs that are set-truncated by the isSet constructor.

## 2.5 Computational UIP in Cubical Agda

Cubical Type Theory (without Glue) with UIP has a simple set model that shows its consistency: types can be interpreted as sets, path types can be interpreted as sub-singleton sets, where hcomp and transp are interpreted as identity functions.

In the definition of $h$-levels, UIP is exactly the assertion that *all types are Sets*, or equivalently, all types have a $h$-level of 2. Therefore, to have UIP in Cubical Type Theory (without Glue Types) is to Set-truncate all types, thus resulting in what we are calling a **$h$-Set Cubical Type Theory**. Since QITs already have the isSet constructor, they *are* already sets; that is why postulating UIP still retains QITs as a feature. The computational way to do this, as opposed to simply postulating UIP as an axiom, is to prove that every type is set-truncated if the base types and higher inductive types are. This is essentially a proof by structural induction on the type derivation, where the assumption of "base types and HITs are set-truncated" act as the base cases, and the preservation of $h$-levels through type formers acts as inductive cases. We detail these proofs in Chapter 4.

---

[4]The HoTT book has an offset of 2 in $h$-levels compared to that in Cubical library: the HoTT book starts from a $h$-level of $-2$ for contractible types instead of 0.

# Chapter 3

# Implementation in Agda

Agda has many extensions and variants which the users can choose from, but some combinations of the extensions are known to be incompatible and can cause inconsistency when used together, either explicitly in a single Agda module, or as a result of importing another module which uses an incompatible option. For example, `--with-K` and `--cubical` cannot exist in the same module since Streicher's K axiom [Str93] is incompatible with univalence. To ensure consistency, Agda enforces checks against some well-known incompatible combinations of extensions in `Agda.Interactions.Options.Base`. As a result, an additional requirement of our implementation of Cubical without Glue is to continue enforcing the consistency of the possible resulting systems.

## 3.1 Cubical without Glue (`--cubical=no-glue`)

The full implementation can be found on https://github.com/agda/agda/pull/7861, and has been merged into mainline Agda as of 16 October 2025 [Tan25b].

Since the Cubical without Glue variant is designed to be compatible with UIP, our implementation ensures that Glue types cannot occur in modules using Cubical Agda without Glue. In particular, there are two key safety requirements:

- Primitives such as Glue Types and their related operations (e.g. the `unglue` function) cannot be defined in the current module.
- Additionally, modules which may contain Glue *cannot* be imported into a module explicitly without Glue, not even if the Glue types are in erased form (i.e. erased when compiled, but might still cause inconsistency).

We implement these requirements by

- Extending the `requireCubical <variant>` effectful check in Agda's Type-Checking Monad (TCM) to ensure a module with `--cubical=no-glue` cannot define the glue primitives, and therefore cannot use them.
- Restricting against the importation of `--cubical` and `--cubical=erased` modules into modules using Cubical without Glue.

Additionally, Cubical without Glue is also a so-called "infective" option: if a module uses Cubical without Glue, every module importing it must activate a Cubical variant. Moreover, when going down the importation chain, the cubical variant must be "increasing" with respect to the following total order:

$$\text{Cubical-compatible} < \text{Cubical without Glue} < \text{Cubical with Erased Glue} < \text{(Full) Cubical}.$$

## 3.2 Primitives in Agda

Primitives in Agda, such as the `hcomp` and `transp` functions of Cubical Agda, are terms with a given type whose computational behaviour is defined using Haskell code. Internal syntax in Agda is defined using an Embedded Domain Specific Language, defined using monads, the `NameT` Reader Monad transformer, arrows and Pi types as monad combinators in order to define terms with named arguments without having to deal with De Bruijn indices. As a proof of concept, I defined UIP as a primitive in https://github.com/SwampertX/agda/tree/cubical-uip which the user at the top level could declare only when the special variant of Cubical Agda without Glue is activated, but no computational rule is defined yet.

# Chapter 4

# Two ways to Square-Fill

Instead of the original UIP statement, we will first show a statement logically equivalent to UIP, which we call the square-filling property SqFill in Section 4.1, then outline our proof by induction on type derivation and discuss the base cases in Section 4.2.

The remainder of the chapter gives proofs of the inductive cases: Section 4.3 gives proofs of the *homogeneous* SqFill property over the type formers Pi, Sigma, Coproducts, and Path Types, and Section 4.4 gives the proofs for the same type formers but for the *heterogeneous* version, SqPFill. Finally, Section 4.5 summarises the complexity of all the proofs in this chapter together with observations on how to possibly simplify them.

Agda proofs (denoted by ↺) are also available on https://swampertx.github.io/hset-cubical/Everything.html.

## 4.1 From UIP to SqFill

> **Remark 4.1.** The diagrams in this chapter consist of
> - nodes, representing objects of a certain type, and
> - directed edges between nodes which represent paths between the corresponding objects.
>
> We represent edges with the following conventions:
> - Directed single edges (⟶) represent generic paths, i.e. proofs of equality.
> - Paths between (single-edged) paths are drawn as directed double edges (⟹).
> - Reflexive proofs of equality are drawn as an undirected double edges (═══).
> - Paths generated by theorems, assumptions, or Kan operations are marked by dashed edges (⇢).
> - Transport-fillers (Proposition 2.1) are drawn as curly edges (⤳).
>
> As an exception that is not a path, a function map is drawn as a directed edge starting with a bar (↦).

### 4.1.1 The Original UIP Statement

> ↺ **Definition 4.2.** (Uniqueness of Identity Proofs):
>
> A type $A$ : Type has the **UIP property**, written as UIP $A$, if the following holds: for any two inhabitants $x\ y : A$, and for any two proofs $p\ q : x \equiv y$ of their equality, the two proofs are equal: $p \equiv q$.

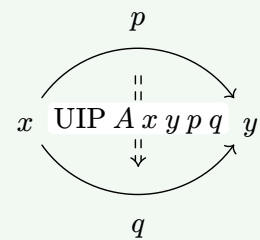

UIP is exactly the assertion that isSet holds for every type, defined in Cubical Agda as below:

```
isSet : (A : Type) → Type
isSet A = (x y : A) (p q : x ≡ y) → p ≡ q

UIP : (A : Type) → isSet A
```

Pictorially, we can think of the points $x\ y : A$ as reflexive sides and extend the picture into a square. Then, the original UIP statement in Definition 4.2 is definitionally equivalent to requiring that any such hollow square (with reflexive sides) can be filled.

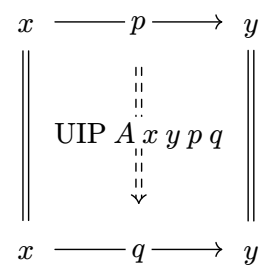

Figure 5: Square with reflexive sides.

**4.1.2 A Generalisation of UIP: The Square-Fill Property**

Interestingly, relaxing the condition of reflexive sides to any paths still results in a statement logically equivalent to UIP. This generalisation was already observed in the Cubical library of Cubical Agda [Agd25], called `isSet'`, states that every hollow square has a filling. This is strictly more general (in terms of usability) than UIP. Since this property plays an important role in the rest of the chapter, we decide to give it a more fitting name: the **square-filling property** (SqFill).

> ↻ **Definition 4.3.** (Square-Filling property): A type $A$ has the square-filling property SqFill $A$ if the following holds: for any hollow square in $A$, that is
> - four corners lu ru ld rd : $A$, and
> - four sides connecting the four corners, namely
>   (i) $l : \text{lu} \equiv_A \text{ld}$
>   (ii) $r : \text{ru} \equiv_A \text{rd}$
>   (iii) $u : \text{lu} \equiv_A \text{ru}$
>   (iv) $d : \text{ld} \equiv_A \text{rd}$
>
> 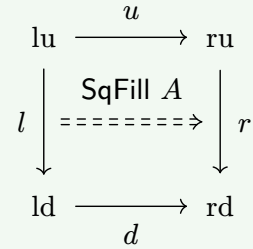
>
> then the square has a filling of type PathP $(\lambda(i : I).u\,i \equiv d\,i)\,l\,r$.

> ↻ **Remark 4.4.** The square produced by SqFill in Definition 4.3 is from left to right. However, we can obtain a square from top to bottom by simply swapping the coordinates $i\,j$, and even swap directions by negating the interval variable: all four possible formulations are logically equivalent.

Let us show that isSet is logically equivalent to SqFill.

> ↻ **Proposition 4.5.** Given any type $A$, the following are logically equivalent:
> (i) Any hollow square in A with reflexive sides has a filling.
> (ii) Any hollow square in A has a filling.

We first need a lemma about giving homogeneous paths instead of heterogeneous paths:

> ↻ **Lemma 4.6.** (Path to PathP): For any line of types $A : I \to \text{Type}$, $x : A\,0$ and $y : A\,1$, if there is a homogeneous path $p : (\text{transport}\,A\,x) \equiv_{(A\,1)} y$ in $A\,1$, then we can get a heterogeneous path PathP $A\,x\,y$.

*Proof.* The proof follows by a hcomp, where the bottom edge (transport-filler $A\,x$) follows from Proposition 2.1.

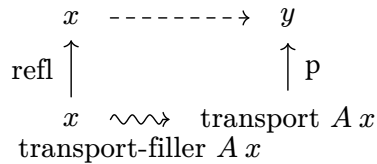

□

Now we can prove the logical equivalence:

*Proof of Proposition 4.5.*

(i) ⇒ (ii): Let a hollow square $l\,r\,u\,d : A$ be given, and we wish to give a path from $l$ to $r$. By Lemma 4.6, It suffices to give a path from transport $(\lambda i \to u\,i \equiv d\,i)\,l$ to $r$. Since transport $(\lambda i \to u\,i \equiv d\,i)\,l$ has the same endpoints as $r$, namely $r\,0$ and $r\,1$, we can simply apply the (ii) assumption.

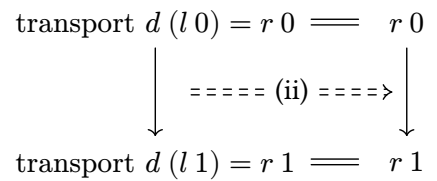

(ii) ⇒ (i): (i) is a special case of (ii) where $l$ and $r$ are reflexive. □

## 4.2 Proving SqFill for type formers

Let us explain what it means to "prove" SqFill for type formers. Our goal is to have a variant of Cubical Type Theory called "h-Set Cubical Type Theory" where **every type has the SqFill property**. We show this by induction on types:

- Base Cases: the base types and the (potentially higher) inductive types have the SqFill property.
- Inductive Cases: type formers preserve the SqFill property.

The usual base types ($\mathbb{0}, \mathbb{1}$) have the SqFill property trivially if the type theory is canonical. The SqFill property for non-HIT inductive types is left as future work, but related proofs about preservation of $h$-levels can be found in the Cubical Library [Agd25]. HITs can be endowed with the SqFill property by adding an isSet constructor. For example, Circle with UIP is simply the following:

```
data Circle : Type where
  base : Circle
  loop : base ≡ base
  uip  : (x y : Circle) (p q : x ≡ y) → p ≡ q
```

For the inductive cases, we leverage a standard result on h-levels in Homotopy Type Theory [Uni13]: in particular, if the constituent types (e.g. $A$, and $B\,a$ for all $a : A$) are h-Sets, the composite types (e.g. $\Pi_{a:A} B\,a$) generated by the type formers are also h-Sets. This result is also proved in Cubical Agda's Cubical library [Agd25] for all h-levels (where applicable), thus for the original UIP statement but not SqFill. While it is possible to conveniently compose such proofs with the SqFill ⇔ UIP proofs to obtain the desired SqFill properties, these abstract proofs are not the most direct, in the sense that it uses more (sometimes also more complex) Kan operations (hcomp, transp) than a direct proof for SqFill.

Since these proofs for SqFill will be implemented into Cubical Agda as computational rules and become part of the new metatheory, the proofs in Sections 4.3 and 4.4 are made as direct as possible to facilitate future analysis of the resulting theory.

## 4.3 Homogeneous Square-Filling: SqFill

This section proves the SqFill property for four type formers: Pi, Sigma, Coproducts, and Path Types, among which the proof for Sigma was exceptionally complicated. See Section 4.5 for a summary.

### 4.3.1 SqFill for Pi Types

> ↻ **Theorem 4.7.** (SqFill-Pi): Given
> - a type $A$ with the SqFill property, and,
> - for every inhabitant $a : A$, a type $B\,a$ with the SqFill property,
>
> then the Pi-type $\Pi(a : A).B\,a$ has the SqFill property.

*Proof.*   To show that $\Pi(a : A).B\,a$ has the SqFill property, we fix a hollow square in the Pi type (Figure 10), and we want to give a filling to the square. That is, fix $i\,j : I$, we want to give a function $f_{ij} : \Pi(a : A).B\,a$ that is equal to $l\,r\,u\,d$ at the boundaries.

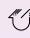

Figure 10: Hollow square of functions given.   Figure 11: Applying $a$ to Figure 10.

*4 Two ways to Square-Fill*Let any $a : A$ be given. Applying it to the given hollow square of functions returns a hollow square in $B\,a$, with sides $\text{lb} := \lambda(j : I).l\,j\,a$, and similarly rb, ub, db as paths in $B$, connecting the corners lub, ldb, rub, rdb $: B\,a$ (Figure 11). Let us define the function $f_{ij}$ as follows:

$$f_{ij} := a \mapsto \mathsf{SqFill}\ (B\,a)\ \text{lb rb ub db}\ i\,j.$$

At the boundaries, $f_{ij}$ is definitionally $l\,r\,u\,d$ since SqFill respects boundaries; thus we are done. □

### 4.3.2 SqFill for Sigma Types

Changing Pi to Sigma in Theorem 4.7 gives us the statement for Sigma:

> ↻ **Theorem 4.8.** (SqFill-Sigma): Given
> (i) a type $A$ with the SqFill property, and
> (ii) for every inhabitant $a : A$, a type $B\,a$ with the SqFill property,
>
> then the Sigma type $\Sigma(a : A).B\,a$ has the SqFill property.

The proof features a proof technique which we will see again in the proofs of SqFill for Coproducts (Theorem 4.10), SqPFill for Pi types (Theorem 4.13) and SqPFill for Path types (Theorem 4.16): what I call the "**transport-fill-align**" method. This method can be applied when the assumption states that a certain type (say $A$) has the SqFill property, but our hollow square has a path-equivalent type ($B$, where $A \equiv B$) but not exactly the right type to use the SqFill assumption. We can then do the following steps to obtain a filling:

- **transport** our hollow square in $B$ to the target type $A$,
- **fill** the resulting hollow square in $A$ by our assumption, and
- **align** the filled square in $A$ back to $B$ with the original boundaries by a comp (or a hcomp if $A \triangleq B$).

The **transport** step in this method is usually enabled by the following lemma, stating that any two types in a square of types are path-equivalent:

> ↻ **Lemma 4.9.** Let $A : I \to I \to \text{Type}$ be a square of types. Then, for any two points in the square, say $A\,i_0\,j_0$ : Type and $A\,i_1\,j_1$ : Type, there is a path $(p : A\,i_0\,j_0 \equiv_{\text{Type}} A\,i_1\,j_1)$ between them.

*Proof.* We can define a "coercion" function on the interval type $I$ from $i_0$ to $i_1$ via $k$ as

$$\text{coe}_k(i_0, i_1) := ((\sim k \vee i_1) \wedge i_0) \vee ((k \vee i_0) \wedge i_1),$$

where $\text{coe}_0(i_0, i_1) = i_0$ and $\text{coe}_1(i_0, i_1) = i_1$. We can define a path between $A\,i_0\,j_0$ and $A\,i_1\,j_1$ by simultaneously going from $i_0$ to $i_1$ and from $j_0$ to $j_1$ using an interval variable $k$:

$$\lambda\,(k : I).A\,(\text{coe}_k(i_0, i_1))\,(\text{coe}_k(j_0, j_1)).$$  □

See also a discussion on eta for $\text{coe}_k$ in Appendix A.1. We are now ready to prove that Sigma preserves SqFill:

*Proof of Theorem 4.8.* To show that $\Sigma(a : A).B\,a$ has the SqFill property, we fix a hollow square in the Sigma type, that is, let a square of dependent pairs in $(a : A)$ and $B\,a$ be given (Figure 13). We want to fill the square: for any coordinate $i\,j : I$, we give a dependent pair of $(a : A)$ and $B\,a$. by their two projections.

The first projection is simple: it suffices to return the filled square on $A$ given by applying the assumption (i) on the first projection of the given hollow square (Figures 14 and 15).

For the second projection, however, it is instead a *heterogeneous* hollow square in a square of types dependent on the first projection. Call the filled square of the first projection (sqa $: \Pi(i\,j : I).A$), then the second projection has a type of $\Pi\,(i\,j : I).B\,(\text{sqa}\,i\,j)$. In the second projection, not all points of the hollow square have the same type; in fact, every point of the hollow square has a *different* type! It is impossible to simply apply assumption (ii): the homogeneous SqFill assumption for $B$.

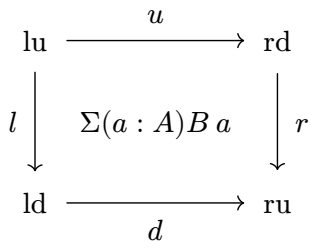

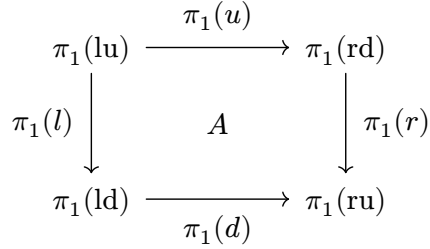

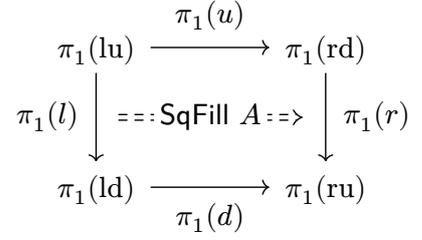

Figure 13: Hollow square in $\Sigma(a : A) B\, a$.

Figure 14: First projection of Figure 13.

Figure 15: Square-filling $A$ by assumption (i).

To obtain a *heterogeneous* filling from a *homogeneous* assumption, we use the "**transport-fill-align**" method on the second projection of the hollow square:

- Fix any $i\ j : I$. By Lemma 4.9, every type of heterogeneous square in $\Pi\ (i'\ j' : I).B\ (\text{sqa}\ i'\ j')$ is pointwise path-equivalent to the *homogeneous* square $\Pi(\_\ \_ : I).B\ (\text{sqa}\ i\ j)$. We would like to transp the heterogeneous square pointwise along these paths of types, with a formula $\varphi$ such that $\varphi = 1 \leftrightarrow (i = i' \wedge j = j')$: the transp is constant at the $(i, j)$-th coordinate. Unfortunately, this is could **not** be done in our De Morgan algebra, which we elaborate in Remark 1.2. We fall back to a **transport** that is nowhere constant, i.e. setting $\varphi := 0$. This is illustrated in Figure 16.

- Then, **fill** in the homogeneous square in $B\ (\text{sqa}\ i\ j)$ by applying the assumption We only require the $(i, j)$-th coordinate of this square.

- After doing the previous two steps for all arbitrary $(i, j)$, we assembled a heterogeneous square of $\Pi\ (i\ j : I).B\ (\text{sqa}\ i\ j)$. However, since we could *not* define the transport step with transp such that the transport is pointwise constant (hence the points are definitionally equal) along the boundaries, we need to finally **align** the boundaries, using a comp to "enforce" the propositional equalities given by transp-fill. This is illustrated in Figure 17.

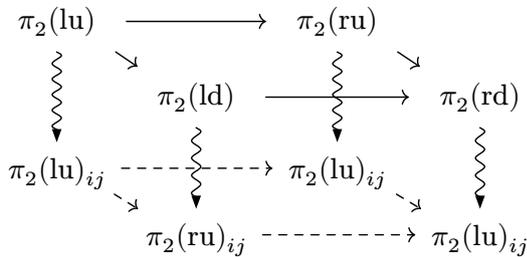

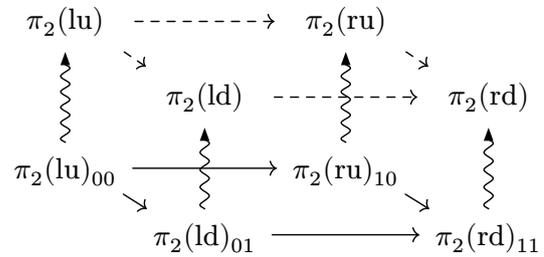

Figure 16: Transporting the heterogeneous hollow (top) square down to $B\ i\ j\ (\text{sqa}\ i\ j)$.

Figure 17: Aligning the sides of the heterogeneous square in $\Pi(i'\ j' : I).B\ (\text{sqa}\ i'\ j')$ by comp. □

### 4.3.3 SqFill for Coproduct Types

> ♡ **Theorem 4.10.** (SqFill-Coproduct): Given types $A$, $B$ with the SqFill property, then the coproduct type $A \uplus B$ has the SqFill property.

*Proof.* The proof is a classic "encode-decode" proof mentioned in the HoTT book [Uni13] and also implemented in the Cubical Agda library [Agd25] for the original h-level formulations and thus about the UIP property instead of SqFill. We adapt the proof for SqFill, which follows identically.

First, define a "code" type, also called Cover in the Cubical Library, which characterises the path space as the coproduct of the constituent path spaces.

We show that Cover is isomorphic to the path space by defining functions encode and decode, then show that they are inverses of each other. Encode is defined by path induction, in which reflexive paths (of inl a or inr b) are mapped identically to their respective path spaces in Cover.

```
Cover : {A B : Type} (c c' : A ⊎ B) → Type     reflCode : {A B : Type} (c : A ⊎ B) → Cover c c
Cover (inl x) (inl y) = x ≡ y                   reflCode (inl x) = refl
Cover (inr x) (inr y) = x ≡ y                   reflCode (inr x) = refl
Cover _ _             = ⊥
                                                encode : {A B : Type} {c c' : A ⊎ B} → c ≡ c' → Cover c c'
                                                encode {c = c} = J (λ c' p → Cover c c') (reflCode c)
```

Decoding is then also identical with respect to $A$ and $B$: inhabitants of Cover, i.e. the coproduct of the path spaces are mapped back to the path space of the coproduct by congruence with respect to inl or inr. Showing decode inverts encode is by another path-induction, since both functions are essentially identity. The base case should have been definitional, however, due to a lack of regularity of transport (and hence $J$, which is defined via transport) of Cubical Type Theory, the propositional equality can be shown using Proposition 2.1.

```
decode : {A B : Type} {c c' : A ⊎ B} → Cover c c' → c ≡ c'
decode {c = inl x} {c' = inl y} = cong inl
decode {c = inr x} {c' = inr y} = cong inr

decodeEncode : {A B : Type} {c c' : A ⊎ B} (p : c ≡ c') → decode (encode p) ≡ p
decodeEncode {c = inl x} = J (λ c' p → decode (encode p) ≡ p) (cong (cong inl) (transportRefl refl))
decodeEncode {c = inr x} = J (λ c' p → decode (encode p) ≡ p) (cong (cong inr) (transportRefl refl))
```

The main proof proceeds as follows. Let a hollow square in the coproduct type $A \uplus B$ be given. Paths in the coproduct type must be either paths in $A$ or paths in $B$ since constructors have the no-confusion property [MGM04]; if we had a path between any inl $a$ and inr $b$, we could directly construct a proof of false. Therefore, it suffices to consider hollow squares of $A \uplus B$ which are entirely inl $A$ or entirely inr $B$. Due to symmetry, let us prove only the case for inl $A$.

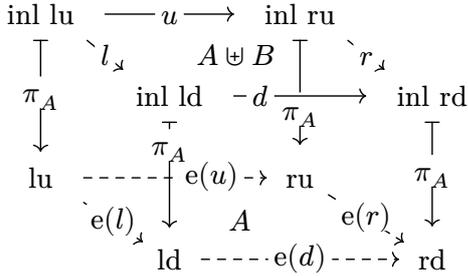

Figure 18: Encoding an inl hollow square from $A \uplus B$ to $A$.

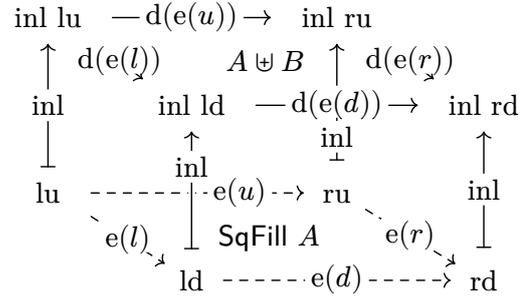

Figure 19: Decoding (applying inl to) the filled square in $A$ back to $A \uplus B$.

We can use encode to get a hollow square in $A$ (Figure 18), and it has a filling given by the SqFill assumption on $A$. Now, we can decode/inl the filled square in $A$ back to $A \uplus B$ (Figure 19), but the boundaries are now of the form $d(e(p))$ instead of just $p$. A final hcomp along the decodeEncode paths as the boundaries (the trapeziums in Figure 20) gives us the desired filling.

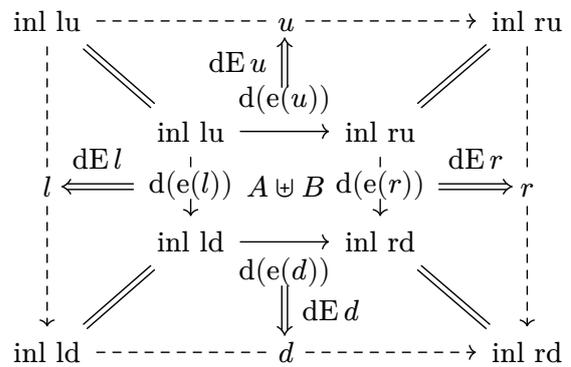

Figure 20: Aligning by hcomp along $d(e(p)) \equiv p$. □

### 4.3.4 SqFill **for Path Types**

> ♡ **Theorem 4.11.** (SqFill-Path): Given a type $A$ with the SqFill property, then for any $a\ b : A$, the path type $a \equiv_A b$ has the SqFill property.

*Proof.* Let a hollow square in the path type $a \equiv_A b$ be given. We want a square of proofs of $a \equiv_A b$ whose edges conform to the given hollow square. Since there are no constraints on the non-edge filling, we can give a constant square any proof of $a \equiv_A b$, for example, just any point of the given hollow square, say the upper-left corner lu : $a \equiv_A b$. Then, it remains to "align" our boundaries with those of the given hollow square using a hcomp operation, then we are done.

In Figure 21, the top face of the cube is the given hollow square, and the bottom face is the constant lu : $a \equiv_A b$ *filled* square. To obtain a filled top face, it remains to hcomp along the side faces, which we must define.

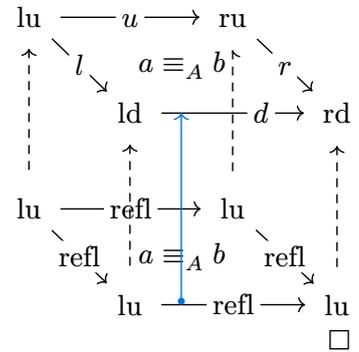

A side face of the cube (Figure 21) is a collection of vertical paths, which are between proofs of $a \equiv_A b$. Since $A$ has SqFill by assumption and is thus a set, by definition, $a \equiv_A b$ is a proposition: all inhabitants of $a \equiv_A b$ are path-equivalent, say lu $\equiv_{a \equiv_A b} (d\ i)$, the line on the face closest to us in Figure 21. We can similarly define all the other faces, and conclude with a final hcomp of the bottom square along these faces defined. □

## 4.4 Heterogeneous Square-Filling: SqPFill

Inspired by the difficulty faced when proving SqFill-Sigma, where the second projection is a heterogeneous square, we generalise the SqFill assumption to the heterogeneous setting. Instead of a single type $A$, we consider the square-filling property in a square of types $A : I \to I \to \text{Type}$.

> ☼ **Definition 4.12.** (SquareP-Filling property): A square of types $A : I \to I \to \text{Type}$ has the heterogeneous square-filling property SqPFill $A$ if the following holds:
>
> For any hollow square in $A : I \to I \to \text{Type}$, that is
> - four corners lu : $A\ 0\ 0$, ru : $A\ 1\ 0$, ld : $A\ 0\ 1$, rd : $A\ 0\ 1$, and
> - four sides connecting the four corners, namely
>   - (i) $l$ : PathP $(\lambda j \to A\ 0\ j)$ lu ld
>   - (ii) $r$ : PathP $(\lambda j \to A\ 1\ j)$ ru rd
>   - (iii) $u$ : PathP $(\lambda i \to A\ i\ 0)$ lu ru
>   - (iv) $d$ : PathP $(\lambda i \to A\ i\ 1)$ ld rd
>
> then the square has a filling: `PathP (λ i → PathP (λ j → A i j) (u i) (d i)) l r`.

Previously, Sigma had a complicated proof because the SqFill assumptions were *homogeneous* while the second projection to be given was *heterogeneous*. By generalising the square-filling statement to the heterogeneous case, the proof for Sigma becomes trivial (i.e. does not need any Kan operation) since the assumption could be applied directly. However, the heterogeneous case of the previously trivial SqFill-Pi now requires a "transport-fill-align". On the other hand, previously simple proofs for coproducts and path types now become trickier since the paths are now heterogeneous.

### 4.4.1 SqPFill **for Pi Types**

> ☼ **Theorem 4.13.** (SqPFill-Pi): Given
> - a square of types $A : I \to I \to \text{Type}$ with the SqPFill property, and
> - a square of types $B : \Pi(i\ j : I).A\ i\ j \to \text{Type}$ dependent on $A$,
>
> such that for every inhabitant $(a : A\ i\ j)$, the type $B\ i\ j\ a$ has the SqPFill property. Then, the square of Pi-types $\lambda(i\ j : I).\Pi(a : A\ i\ j).B\ a$ has the SqPFill property.

The proof uses the **transport-fill-align** method.

*Proof of Theorem 4.13.* Let a hollow square in the square of Pi types be given, and we want to fill the square. Let us fix $i\ j : I$. We want to construct a dependent function $f_{ij} : \Pi(a : A\ i\ j).B\ i\ j\ a$. Intuitively, we would like to define the function $f_{ij}$ by sending $(a : A\ i\ j)$ to the $(i, j)$-th coordinate of the square filled from the hollow square in $B\ i\ j\ a$, formed from $\text{lb} := \lambda\ (j : I).l\ j\ a$, and similarly defined rb, ub, db.

$$f_{ij} \stackrel{?}{:=} (a : A\ i\ j) \mapsto \mathsf{SqPFill}\ (\lambda\ i\ j.B\ i\ j\ a)\ \text{lb rb ub db}\ i\ j,$$

However, this naive attempt now does not type check: previously in the *homogeneous* case of Theorem 4.7, the points on the boundaries were functions of type $\Pi(a : A).B\ a$ with the same domain and thus can be applied to the argument $a : A$ directly. However, here they have domains that vary along the coordinates. For example, the left edge

$$l : \mathrm{PathP}\ (\lambda\ (j' : I).A\ 0\ j')\ \text{lu ld}$$

has domain types varying from $A\ 0\ 0$ to $A\ 0\ 1$. Therefore, we must first transport the argument $a : A\ i\ j$ to the corresponding types (e.g. $A\ 0\ 0$) of the function domains before applying, made possible by Lemma 4.9.

```
spread : {i j : I} (a : A i j) (i' j' : I) → A i' j'
spread a i' j' = transport (λ k → A (icoe i i' k) (icoe j j' k)) a
```

The proof uses the **transport-fill-align** method as follows:

- Fix $(i\ j : I)$. We apply the hollow square of functions to the square $\lambda(i'\ j' : I).\mathsf{spread}\ a\ i'\ j'$ and obtain a hollow, heterogeneous square in the square of types $\lambda\ (i'\ j' : I).B\ i'\ j'\ (\mathsf{spread}\ a\ i'\ j')$. We denote $(\mathsf{spread}\ a\ i\ j)$ as $a_{ij}$ in Figure 23. Here, spread suffers similarly from the inability to be constant when $i' = i$ and $j' = j$, which cannot be described as the $\varphi$ in a transp, and is thus defined as a transport. The discussion in Appendix A.2 applies here too.

- Now we can apply assumption (ii) on this heterogeneous hollow square and get a (heterogeneous) filling. Its $i' = i, j' = j$ coordinate of type $B\ i\ j\ (\mathsf{spread}\ a\ i\ j)$ is what we needed, given that we have $\mathsf{spread}\ a\ i\ j \equiv a$, which follows from Proposition 2.1.

- after doing the previous two steps for all arbitrary $i\ j : I$, now we have a (heterogeneous) filling of the given square, but in the slightly different type $\Pi(i\ j : I).B\ i\ j\ (\mathsf{spread}\ a\ i\ j)$ instead of $\Pi(i\ j : I).B\ i\ j\ a$, which are *not* definitionally equal due to the first transport step. Therefore, a comp along the transport-fillers given in Proposition 2.1 is needed to get back to the right type and realign the boundaries (Figure 23).

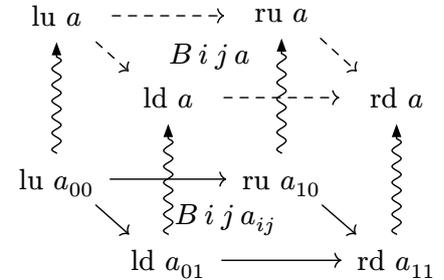

Figure 23: Aligning the sides by comp.

### 4.4.2 SqPFill for Sigma Types

> **Theorem 4.14.** (SqPFill-Sigma): Given
> - a square of types $A : I \to I \to \mathrm{Type}$ with the SqPFill property, and
> - a square of types $B : (i\ j : I) \to A\ i\ j \to \mathrm{Type}$ dependent on $A$
>
> such that for every inhabitant $a : (i\ j : I) \to A\ i\ j$, the type $B\ i\ j\ (a\ i\ j)$ has the SqPFill property. Then, the square of Sigma-types $\lambda(i\ j : I).\Sigma(a : A\ i\ j).B\ a$ has the SqPFill property.

*Proof.* To show that $\Sigma(a : A\ i\ j).B\ a$ has the SqPFill property, we fix a hollow square in the Sigma type, that is, let a square of dependent pairs in $(a : A\ i\ j)$ and $B\ i\ j\ a$ be given. We want to give a filling to the square: for any coordinate $i\ j : I$, we want to provide a dependent pair of $(a : A\ i\ j)$ and $B\ i\ j\ a$. We define it in two steps: the first and the second projections.

The first projection is simple: it suffices to return the filled square on A given by applying the assumption (i) on the first projection of the hollow square. So is the second projection: previously we struggled since our SqFill assumption was homogeneous, which was insufficient to fill a *heterogeneous* square. Now we have a heterogeneous SqPFill assumption on $B$, which we can apply directly to obtain the second projection, and no Kan operation was needed. □

### 4.4.3 SqPFill for Coproduct Types

The coproduct for two squares of types $A, B : I \to I \to \text{Type}$ is a square of types, where an inhabitant $c : (A + B) \, i \, j$ has the form of either
- $c = \text{inl } a$ for some $a : A \, i \, j$ or
- $c = \text{inr } b$ for some $b : B \, i \, j$.

> ↻ **Theorem 4.15.** (SqPFill-Coproduct): Given two squares of types $A, B : I \to I \to \text{Type}$ with the SqPFill property, then the coproduct type $A \uplus B : I \to I \to \text{Type}$ has the SqPFill property.

The proof is by the encode-decode method almost identical that of Theorem 4.10, modulo the heterogeneity: the filled square in $A$ or $B$ now needs a comp instead of a hcomp to align its boundaries. We omit the proof here and invite interested readers to read the proof in Cubical Agda.

### 4.4.4 SqPFill for Path Types

> ↻ **Theorem 4.16.** (SqPFill-Path): Given a square of types $A : I \to I \to \text{Type}$ with the SqPFill property, then for any $a \, b : \Pi(i \, j : I).A \, i \, j$, the square of path types $\lambda \, (i \, j : I).a \, i \, j \equiv_{A \, i \, j} b \, i \, j$ has the SqPFill property.

To simplify the "transport-fill-align" proof later, here is a generic lemma:

> ↻ **Lemma 4.17.** (QuadFill): Let $A : I \to I \to \text{Type}$ be a square with the SqPFill property. Then for any four corners in $A$, the quadrilateral given by
>
> - four corners:
>   - lu : $A \, i_{\text{lu}} \, j_{\text{lu}}$
>   - ru : $A \, i_{\text{ru}} \, j_{\text{ru}}$
>   - ld : $A \, i_{\text{ld}} \, j_{\text{ld}}$
>   - rd : $A \, i_{\text{rd}} \, j_{\text{rd}}$, and
> - four sides:
>   - $l$ : PathP $(\lambda(k : I).A \, (\text{coe}_k(i_{\text{lu}}, i_{\text{ld}})) \, (\text{coe}_k(j_{\text{lu}}, j_{\text{ld}})))$ lu ld
>   - $r$ : PathP $(\lambda(k : I).A \, (\text{coe}_k(i_{\text{ru}}, i_{\text{rd}})) \, (\text{coe}_k(j_{\text{ru}}, j_{\text{rd}})))$ ru rd
>   - $u$ : PathP $(\lambda(k : I).A \, (\text{coe}_k(i_{\text{lu}}, i_{\text{ru}})) \, (\text{coe}_k(j_{\text{lu}}, j_{\text{ru}})))$ lu ru
>   - $d$ : PathP $(\lambda(k : I).A \, (\text{coe}_k(i_{\text{ld}}, i_{\text{rd}})) \, (\text{coe}_k(j_{\text{ld}}, j_{\text{rd}})))$ ld rd,
>
> has a filling.

*Proof.* The proof follows by a "**transport-fill-align**":
- First transport the hollow quadrilateral to the square of types in $A$ as defined in the SqPFill statement: top left in $A \, 0 \, 0$ and bottom right in $A \, 1 \, 1$, along the paths of types given pointwise by Lemma 4.9.
- Then, fill the square by the SqPFill assumption of $A$.
- Finally, comp the resulting filled square back to the quadrilateral, with the sides given by transp-fill. □

Now we are ready to prove that Path types preserve SqPFill. The proof proceeds in exactly the same way as the homogeneous case Theorem 4.11, except that the isProp fact used previously is replaced by our generic QuadFill lemma above.

*Proof of Theorem 4.16.* Let a hollow square in the path types $\lambda(i \, j : I).a \, i \, j \equiv_{A \, i \, j} b \, i \, j$ be given. We want a square of proofs of $a \, i \, j \equiv_A b \, i \, j$ whose edges conform to the given hollow square. Since there are no constraints on the non-edge filling, we can give a constant square any proof of $a \, i \, j \equiv_{A \, i \, j} b \, i \, j$, for example, the upper-left corner lu. Then, it remains to "align" our boundaries with those of the given hollow square using a comp operation, then we are done.

In Figure 24, the top face of the cube is the given hollow square in $a\ i\ j \equiv_{A\ i\ j} b\ i\ j$, and the bottom face is the constant lu : $a\ 0\ 0 \equiv_{A\ 0\ 0} b\ 0\ 0$ (filled) square.

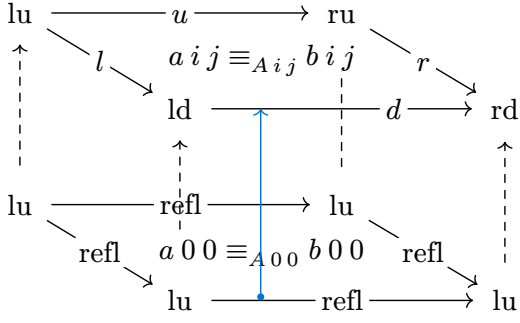
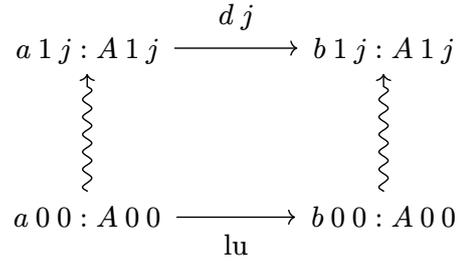

Figure 24: comp-ing to obtain the top face.

Figure 25: The blue line on the side of Figure 24 as a square.

If we can define the side faces, we can conclude by applying a comp on the filled bottom square. The sides were previously obtained by isProp $A$, but now our paths are heterogeneous: the line on the side face closest to us in Figure 24 has a type of $a\ 0\ 0 \equiv_{A\ 0\ 0} b\ 0\ 0$ at the bottom and type $a\ 1\ j \equiv_{A\ 1\ j} b\ 1\ j$ on top. Since its endpoints are inhabitants of path types themselves, this path is essentially a heterogeneous square in $A\ i\ j$, which we can see by expanding its endpoints into Figure 25, with the vertical sides given by transp-fill. Now it suffices to find a filling for the square in Figure 25, for which we can apply Lemma 4.17.

Once the side faces of Figure 24 are defined, the proof concludes with a final comp. □

## 4.5 Summary of SqFill and SqPFill Proofs

The complexity of various proofs of SqFill and SqPFill, which are potential computational UIP rules, are summarised in the following table. We explain some possibilities of simplification as well.

|  | SqFill (*homogeneous* Square-Filling) | SqPFill (*heterogeneous* Square-Filling) |
|---|---|---|
| Pi | Trivial (no Kan operations) | Complicated: transport-fill-align*† |
| Sigma | Complicated: transport-fill-align* | Trivial (no Kan operations) |
| Coproducts | Standard encode-decode proof ($J$, irregularity‡, hcomp) | Standard encode-decode proof ($J$, irregularity‡, comp) |
| Path Types | Simple (a single hcomp) | Complicated: transport-fill-align*^ |

\* : If we could define an equality function on $I$, we could have skipped the final "align" step as well as the definition of the faces for the final hcomp or comp. See a discussion in Appendix A.2.

^ : A stronger induction hypothesis stating that *any* square in the cube of types $A : I \to I \to I \to$ Type had a filling (instead of just the squares $A_k := (\lambda(i\ j : I).A\ i\ j\ k$ for any $k$) could replace the QuadFill lemma to simplify the proof. The QuadFill lemma essentially derives a stronger hypothesis from the existing one.

† : The transport step here needed a definition of coercion of path variables $\text{coe}_k$ that is definitionally constant when the endpoints are equal: $\text{coe}_k(i, i) = i$. We give such an implementation in Appendix A.1.

‡ : The proof here can be slightly simplified if transport (in which $J$ was defined) had the so-called regularity property: transporting along a reflexive path is definitionally the identity function, which does *not* hold in Cubical Type Theory [Swa18].

# Chapter 5

# Future and Related Work

In conclusion, I extended Cubical Agda with a `--cubical=no-glue` option, as well as detailed proofs of SqFill and SqPFill, the square-filling properties equivalent to UIP, in homogeneous and heterogeneous forms.

## 5.1 Future Work

### 5.1.1 What Makes a Computation Rule *Good*?

The simplest computation rules are free of Kan operations and follow directly from their assumptions. However, if Kan operations are necessary, as with most of the proofs of Chapter 4 (the only exceptions being Theorem 4.7 and Theorem 4.14), the criteria to compare which computation rule is better are unclear. For example, our proof of SqFill-Sigma uses a "transport-fill-align" strategy to give a filling directly, while the proof of isSet-Sigma in the Cubical Library (`Cubical.Foundations.HLevels`) uses facts such as the isomorphism between PathPs and Paths (a stronger version of Lemma 4.6), $h$-levels of paths, as well as the uniqueness of hcomp. Our approach is more direct and self-contained but slightly more technical, while the latter is "easier to write" with a higher level of abstraction, but unclear if its dependent proofs and their respective Kan operations will make for a "better" computational rule; the latter also requires more Kan operations in a crude comparison. We leave the evaluation of the "goodness" of these computational rules as future work, to be discussed with other researchers and Agda developers.

### 5.1.2 More Type Formers and Implementation

The preservation of SqFill and SqPFill by inductive types (or $W$-types) and heterogeneous path types (which are provable but slightly more technical than Path types), as well as the computational behaviour for the universe of all types are left as future work, along with the implementation of all the computation rules into a new `--cubical=uip` variant.

### 5.1.3 Normalisation and Canonicity

Our resulting theory of (Cubical Type Theory - Glue + computational UIP) is consistent by a set model sketched in Section 2.5 or possibly via a translation to XTT [SAG22], the next step would be to show its normalisation and canonicity. Although we have not been able to prove them, since our computational rules are structural with respect to type constructors, this suggests that these properties may be proven similarly to existing proofs for regular Cubical Type Theory [Hub16, SA21].

### 5.1.4 Compatibility with Axiom K

Since our `--cubical=no-glue` is compatible with UIP, whether pattern-matching on Identity types (not to be confused with Path Types) [Swa18] using the logically equivalent Streicher's axiom K [Str93] should be allowed in Cubical without Glue remains an open question. In this case, the `--cubical=compatible` flag which all Agda Cubical variants (including `--cubical=no-glue`) depend on should be restructured to *not* imply `--without-K`.

## 5.2 Related work

### 5.2.1 XTT

XTT [SAG22] is a Cubical Type Theory (without Glue Types) with definitional UIP: two paths are *definitionally* equal if they have the same endpoints, while our computation rules provide only *propositional* equality. Furthermore, it is also possible to show the consistency of our theory by a translation into XTT.

### 5.2.2 Setoid Type Theory

Setoid models of Type Theory [Alt99, Alt+19, Hof95] add functional extensionality, propositional extensionality, and quotient types to intensional type theory, and also have more observational properties. For example, in Setoid Type Theory, equality between pairs is *definitionally equal* to the pointwise equalities of the first and second components, but only an isomorphism in Cubical Type Theories.

# Appendix

## A.1 Coercion with eta on the Interval Type $I$

> ↻ **Remark 1.1.** ($\text{coe}_k$ with $\eta$): The coercion operator $\text{coe}_k(i_0, i_1)$ used in Theorem 4.13 must be constant when the arguments are equal, i.e. $\text{coe}_k(i, i) = i$ definitionally. However, in the usual "if-then-else" encoding of Boolean Algebra, the definition
>
> $$\text{coe}'_k(i, j) := (\sim k \wedge i) \vee (k \wedge i)$$
>
> does not have this property:
>
> $$\text{coe}'_k(i, i) \triangleq (\sim k \wedge i) \vee (k \wedge i) = i \wedge (k \vee \sim k) \neq i \wedge 1 = i$$
>
> This is not definitionally $i$ unless excluded middle in $I$ holds, e.g. $I$ is a Boolean algebra. Fortunately, our definition of $\text{coe}_k$ in Lemma 4.9 has this property:
>
> $$\text{coe}_k(i_0, i_1) := ((\sim k \vee i_1) \wedge i_0) \vee ((k \vee i_1) \wedge i_0) \text{ where}$$
>
> $$\begin{aligned}
\text{coe}_k(i, i) &\triangleq ((\sim k \vee i) \wedge i) \vee ((k \vee i) \wedge i) \\
&= ((\sim k \vee i) \vee (k \vee i)) \wedge i && \text{(distributivity)} \\
&= (k \vee \sim k \vee i) \wedge i \\
&= (i \vee (0 \wedge (k \vee \sim k))) = i \vee 0 = i.
\end{aligned}$$

Our formulation essentially eliminates the excluded middle expression by syntactically enriching the $k$ and $\sim k$ expressions with an $i$ that can be "absorbed" due to distributivity of De Morgan Algebras. In fact, enriching only one of the two conjunctions in $\text{coe}'_k$ also yields the desired property.

## A.2 Equality Function in De Morgan Algebra

> **Remark 1.2.** The **transport** step in Theorem 4.8 needed a $\varphi$ whose set of solutions when $\varphi = 1$ corresponds exactly to the condition $i = i' \wedge j = j'$; in this case, the **align** step along, for example lu $a_{00}$ to lu $a$ would have been unnecessary since
>
> $$a_{ij} = \text{spread } a(i' = i)(j' = j) \triangleq a$$
>
> would hold definitionally. If the interval type $I$ had been a Boolean algebra, we could have defined $\varphi$ using the rather familiar equality function:
>
> $$\text{eq}(i, j) := (i \wedge j) \vee (\sim i \wedge \sim j)$$
>
> which does not work in a De Morgan algebra:
>
> $$\text{eq}(i, i) = i \vee \sim i$$
>
> which is not definitionally equal to $1 : I$ in general. However, if such a definition existed, the "align" step of the previous proof (of Theorem 4.8) would no longer be needed.